\documentclass[journal=jctcce,manuscript=article,layout=twocolumn]{achemso}
\usepackage{graphicx}
\usepackage{dcolumn}
\usepackage{bm}
\usepackage{graphicx,amsmath}
\usepackage{tikz}
\usepackage[version=4]{mhchem} 
\usetikzlibrary{quantikz}
\usepackage{subfigure}

\usepackage{algpseudocode}
\usepackage{algorithm}

\usepackage{makecell}
\usepackage{physics}
\usepackage{amssymb}
\usepackage{placeins}

\makeatletter
\newcommand{\citenumtext}[1]{%
  \begingroup
  \let\@cite\NAT@citenum
  \citenum{#1}%
  \endgroup
}
\makeatother

\title{
Initial-State Preparation for Strongly Correlated Molecules from Matrix Product States: Benchmarking and Environment-Aware Divide-and-Conquer Compilation
}
\author{Haiqi Liu}
\affiliation{Key Laboratory of Theoretical and Computational Photochemistry, Ministry of Education, College of Chemistry, Beijing Normal University, Beijing, 100875, China}
\alsoaffiliation{Institute for Advanced Study, Beijing Normal University, Beijing, 100875, China}

\author{Longfei Chang}
\affiliation{Key Laboratory of Theoretical and Computational Photochemistry, Ministry of Education, College of Chemistry, Beijing Normal University, Beijing, 100875, China}
\alsoaffiliation{Institute for Advanced Study, Beijing Normal University, Beijing, 100875, China}

\author{Zhendong Li}\email{zhendongli@bnu.edu.cn}
\affiliation{Key Laboratory of Theoretical and Computational Photochemistry, Ministry of Education, College of Chemistry, Beijing Normal University, Beijing, 100875, China}
\alsoaffiliation{Institute for Advanced Study, Beijing Normal University, Beijing, 100875, China}

\author{Wei-Hai Fang}
\affiliation{Key Laboratory of Theoretical and Computational Photochemistry, Ministry of Education, College of Chemistry, Beijing Normal University, Beijing, 100875, China}
\alsoaffiliation{Institute for Advanced Study, Beijing Normal University, Beijing, 100875, China}

\begin{document}

\begin{tocentry}

\includegraphics[width=8.25cm]{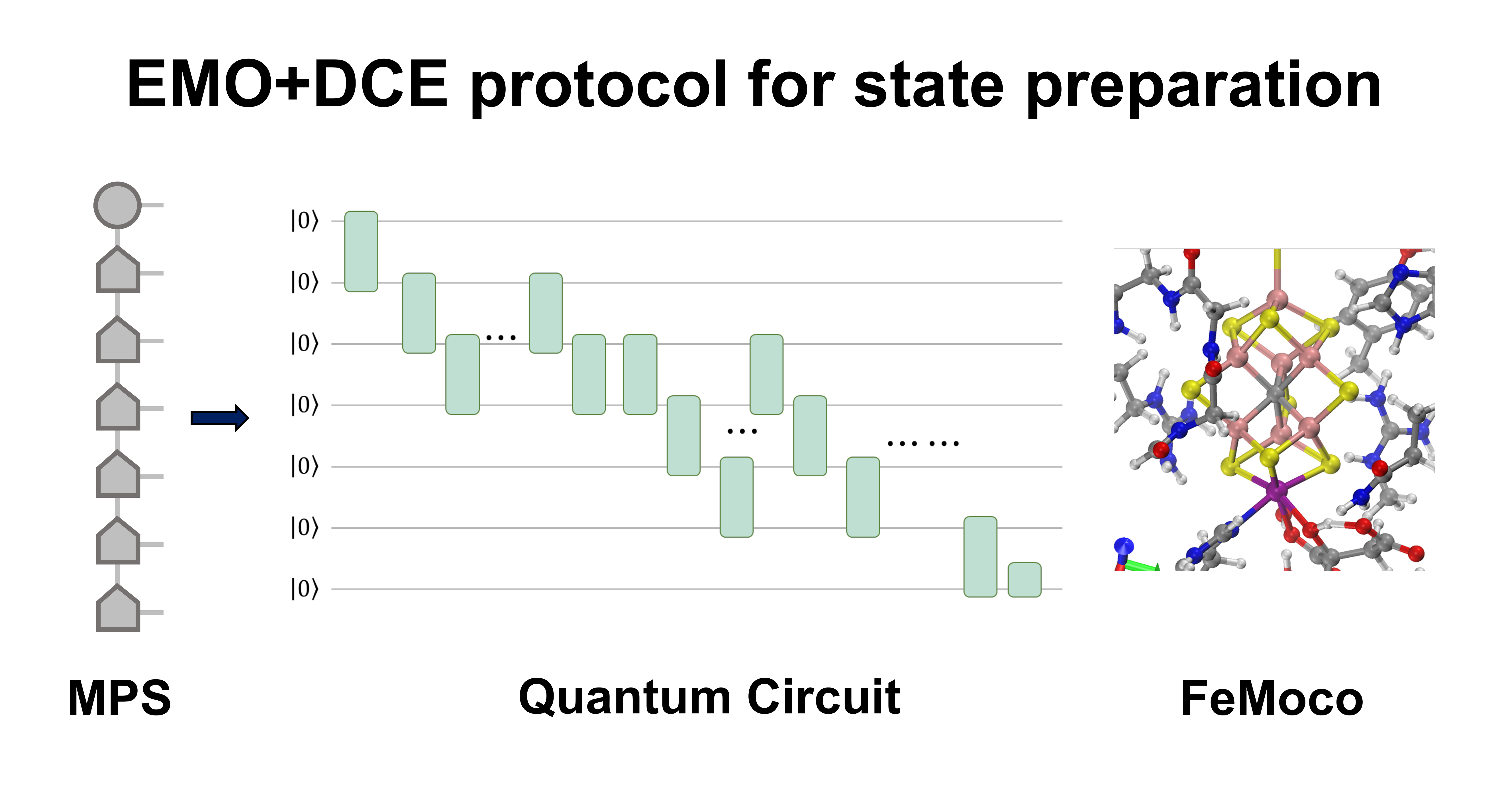}

\end{tocentry}

\begin{abstract}
Compiling classically optimized matrix product states (MPSs) into quantum circuits is a promising route to initial-state preparation for strongly correlated molecules, but the accuracy, gate cost, and classical scalability of existing MPS compilation methods have not been systematically assessed. Here, we benchmark representative analytical, variational, and hybrid MPS compilation methods for strongly correlated systems including hydrogen chains and iron-sulfur clusters. We compare state infidelity, CNOT gate count, classical runtime, and parallel scalability across different system sizes and bond dimensions. We also introduce a divide-and-conquer environment-aware compilation algorithm (DCE), which optimizes each local isometry approximation according to its contribution to the prepared-state fidelity. The results reveal complementary operating regimes: direct variational compilation methods are gate-efficient for some smaller molecular systems with $O(10)$ qubits and $O(10)$ bond dimensions or moderate accuracy targets with infidelity of order $10^{-1}$, whereas divide-and-conquer methods scale more favorably for larger systems and relatively higher bond dimensions of $O(10\text{-}10^2)$, particularly when target infidelities below $10^{-1}$ are required. Within the latter family, environment-aware optimization improves accuracy over unitary-distance-based compilation while retaining task-level parallelism. 
For the FeMo-cofactor (FeMoco), we find the DCE algorithm together with the MPS obtained in the entanglement-minimized orbitals (EMOs) provides a resource-efficient route to initial-state preparation. Our results provide practical guidance for selecting and developing MPS compilation methods for strongly correlated molecular simulations. 
\end{abstract}

\maketitle

\section{Introduction}
Quantum chemistry is fundamentally concerned with solving the electronic structure problem for many-electron systems. As the numbers of electrons and orbitals increase, the rapid growth of the many-electron Hilbert space makes exact solutions of the electronic Schr\"{o}dinger equation prohibitively expensive beyond small systems.~\cite{CaoQChem2019,McArdleQCChem} Although approximate electronic structure methods have greatly extended the range of accessible systems, achieving high accuracy for strongly correlated problems within tractable computational resources remains a major challenge.~\cite{P-3,ChanQA2024}

Quantum computers offer an alternative framework in which many-electron wavefunctions are represented and manipulated directly as quantum states. On future fault-tolerant quantum computers, quantum phase estimation\cite{kitaev1995quantum} (QPE) is expected to be a central algorithm for resolving the eigenspectrum of molecular Hamiltonians. Since the success probability of QPE is proportional to the squared overlap of the initial state with the target eigenstate, a smaller overlap results in increased runtime and quantum resource requirements.\cite{LouvetQPE2026,ChanQA2024} Efficient preparation of high-quality initial states is thus essential for the practical application of QPE to quantum chemistry.

A range of strategies has been explored for preparing such initial states, with representative examples including adiabatic evolution,\cite{Albashadiabatic,Aspuruadiabic} dissipative state preparation,\cite{Zoller_dissipatively,Cirac_dissipatively} and variational quantum algorithms.\cite{Liu2019,Fan2023,GerberQMERA2024} These strategies, however, carry out a substantial part of the preparation procedure on the quantum processor and may require long coherent evolutions or repeated circuit evaluations. An alternative is to construct and optimize an approximate wavefunction on a classical computer and subsequently compile it into a quantum circuit, thereby shifting much of the computational effort to classical preprocessing and reducing the burden of state preparation on the quantum processor.\cite{Rudolph2023}

Encoding approximate wavefunctions in the sum of Slater determinants form provides a practical route to initial state preparation.\cite{TubmanSOS2018,FomichevSOS2024} However, strongly correlated systems may require many determinants for an accurate description, limiting the range of states that can be implemented efficiently in the sum of Slater determinants form.
An alternative strategy is to first obtain an accurate tensor-network approximation to the target state and then compile it into a quantum circuit.\cite{Jobst2024,Haghshenas2022,SugawaraTTN2025,MaccormackPEPs2021} Among tensor-network representations, matrix product states\cite{schollwock2011density} (MPSs) are the most widely used in quantum chemistry\cite{ChanDMRG2011} and, with a suitable bond dimension, can provide high-quality approximations to the ground states of many strongly correlated systems. Their compilation into quantum circuits may be either exact or approximate. In the exact mapping, an MPS is represented by a single sequential layer of local multi-qubit unitaries, which can subsequently be decomposed exactly into native gates supported by quantum hardware,\cite{schon2005,bellante2026,huggins2025,rupprecht2026,Berryrapid2025} while approximate MPS compilation methods relax the requirement of an exact circuit representation.
They address different quantum-resource regimes. In large-scale fault-tolerant phase estimation, exact preparation of a high-bond-dimension MPS can be justified when Hamiltonian simulation overwhelmingly dominates the total non-Clifford cost. For example, recent resource estimates for the FeMo-cofactor (FeMoco) indicate that an MPS with a bond dimension of several thousand can provide a high-overlap initial state while its exact preparation remains a relatively small contribution to the overall phase-estimation cost.\cite{Berryrapid2025} In more resource-constrained settings, however, exact isometry synthesis may require excessive gate counts, circuit depth, or ancillary qubits. 

Approximate MPS compilation methods then provide an alternative by trading a controlled reduction in state fidelity for a potentially substantial reduction in circuit resources.
They can be broadly classified into three categories: 
analytical, variational, and hybrid methods.
\cite{Ciraclogdepth2024,SmithGirvin2024,BohunVlayer2026,WeiPScircuit2026,mingare2026,GuoImprovedMPS2025}
The matrix product disentangler\cite{Ran2020} (MPD) is the representative analytical algorithm, exploiting the exact representation of an MPS with bond dimension \(\chi=2\) by a single staircase circuit of one- and two-qubit gates. However, disentanglers constructed solely from low rank approximations without further optimization can be ineffective for many target MPSs, yielding little improvement from additional layers.

Variational compilation methods have therefore received broader attention. They construct higher fidelity circuits mainly by maximizing the overlap with the target state or the weight of leading Schmidt coefficients.
\cite{SchuchBrickwall2026,Pollmann2021,CrainADAPTAQC2026,Rudolph2024,Schmelcher2026,ShirakawaAQCE2024,ZhukAQCTensor2025,MartinQMPSOevo2024,Chai2026,dborin2021} However, as the number of circuit layers increases, the overlap between the prepared state and the target MPS often saturates at a value that tends to be lower for larger and more complex systems, thereby restricting the practical applicability of these methods.\cite{Rudolph2024,WangSSO2025,ShirakawaAQCE2024,Torre2024,RanADQC2021,MelnikovRie2023,GreenWang2025,iqbal2022} Moreover, many algorithms update gates one by one or layers sequentially, limiting parallelization and increasing the wall time as the system size and the number of circuit layers grow.\cite{Rudolph2024,gibbs2026,Torre2024,Pollmann2021}

Hybrid methods combine exact MPS decomposition with variational approximation of the resulting multi-qubit operations, without showing the overlap saturation observed for direct circuit optimization in the benchmarks considered. Rungger et al.\cite{Rungger2025} sequentially compile the intermediate states defined by the exact MPS decomposition, using the accumulated approximate state from the preceding step as the input to the next step. This sequential dependence limits parallelism, while the number of required compilation steps increases with system size. Su et al.\cite{YaoTingSu2025} use a divide-and-conquer method that approximates each multi-qubit gate independently by minimizing its Hilbert-Schmidt distance from a layered ansatz of two-qubit gates. Independent compilation of the individual gates permits parallel execution and may therefore mitigate the growth of classical wall-clock time with system size. 

In this work, we address the approximate encoding problem through two complementary contributions. 
First, since available analytical, variational, and hybrid MPS decomposition methods have not been systematically benchmarked for chemical systems before, we benchmark their performances on strongly correlated systems including hydrogen chains and iron-sulfur clusters. 
Second, we introduce a divide-and-conquer environment-aware compilation algorithm (DCE), which is more suitable for $O(10\text{-}10^2)$ bond dimensions and large number of qubits. This target regime is motivated by the recently introduced entanglement-minimized orbitals (EMOs)\cite{EMO}, which can make the wavefunction highly compact through entanglement minimization. Consequently, MPS with modest bond dimensions can be used as good initial state for QPE.
The DCE algorithm independently approximates the exact MPS isometries by optimizing the corresponding approximation’s contributions to the prepared-state fidelity. It expands the accessible fidelity–circuit-cost frontier for moderate effective bond dimensions, while retaining task-level parallelism suitable for large systems. The benchmark comparison among different methods jointly considers state infidelity, CNOT gate count, classical compilation time, system-size dependence, and parallel scalability, thereby identifying the resource regimes in which different MPS compilation methods are advantageous. 
An open-source Python-based implementation of different approximate compilation methods is available in the \texttt{mps2circuit} module\cite{FOCUS-MPS2Circuit} of the \textsc{FOCUS} package\cite{li2021expressibility,EMO}.

The remaining parts of this paper is organized as follows:
Sec. \ref{sec:theory} summarizes different MPS compilation methods
for compiling MPS into quantum circuits.
Sec. \ref{sec:results} include benchmarks
for hydrogen chains and iron-sulfur clusters.
Conclusion and outlook are drawn in Sec. \ref{sec:conclusion}.

\section{MPS compilation methods}\label{sec:theory}
\subsection{Matrix product state (MPS) representation}

\begin{figure*}[t]
    \centering
    \includegraphics[width=0.9\textwidth]{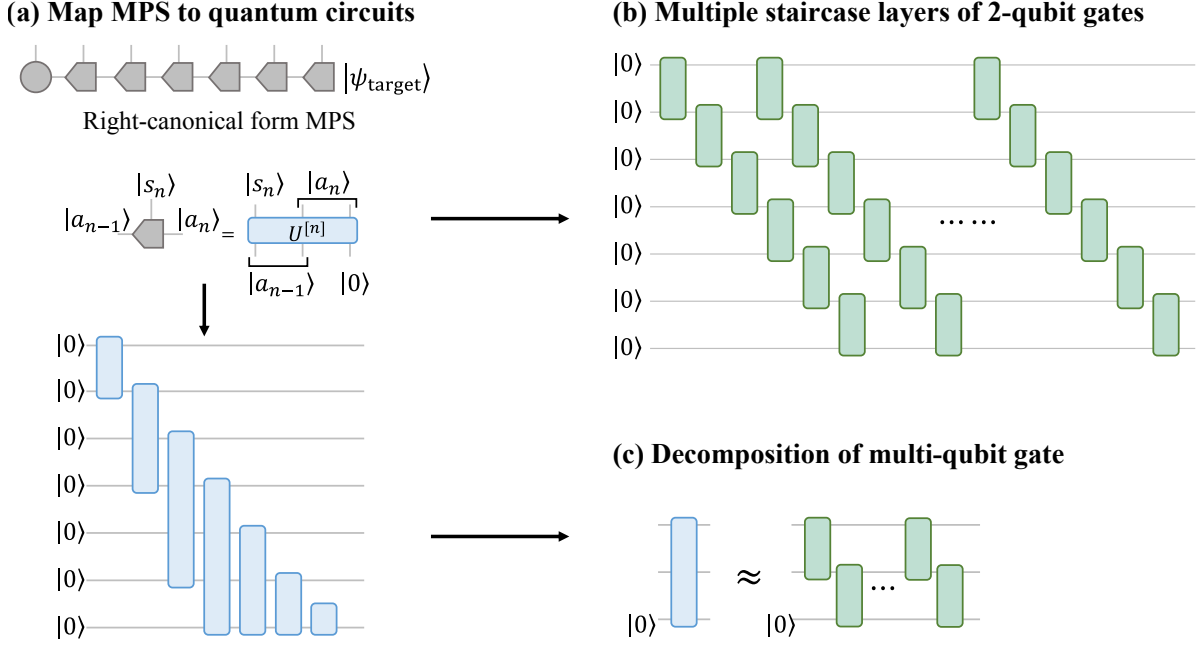}
    \caption{MPS decomposition methods. (a) Exact decomposition of a right-canonical MPS into sequential multi-qubit gates. (b) Approximate decomposition of an MPS into staircase layers of two-qubit gates. (c) Approximate decomposition of each multi-qubit gate obtained in (a) into multiple two-qubit gates.}
    \label{fig:circuit}
\end{figure*}
\begin{figure*}[t]
    \centering
    \includegraphics[width=0.98\textwidth]{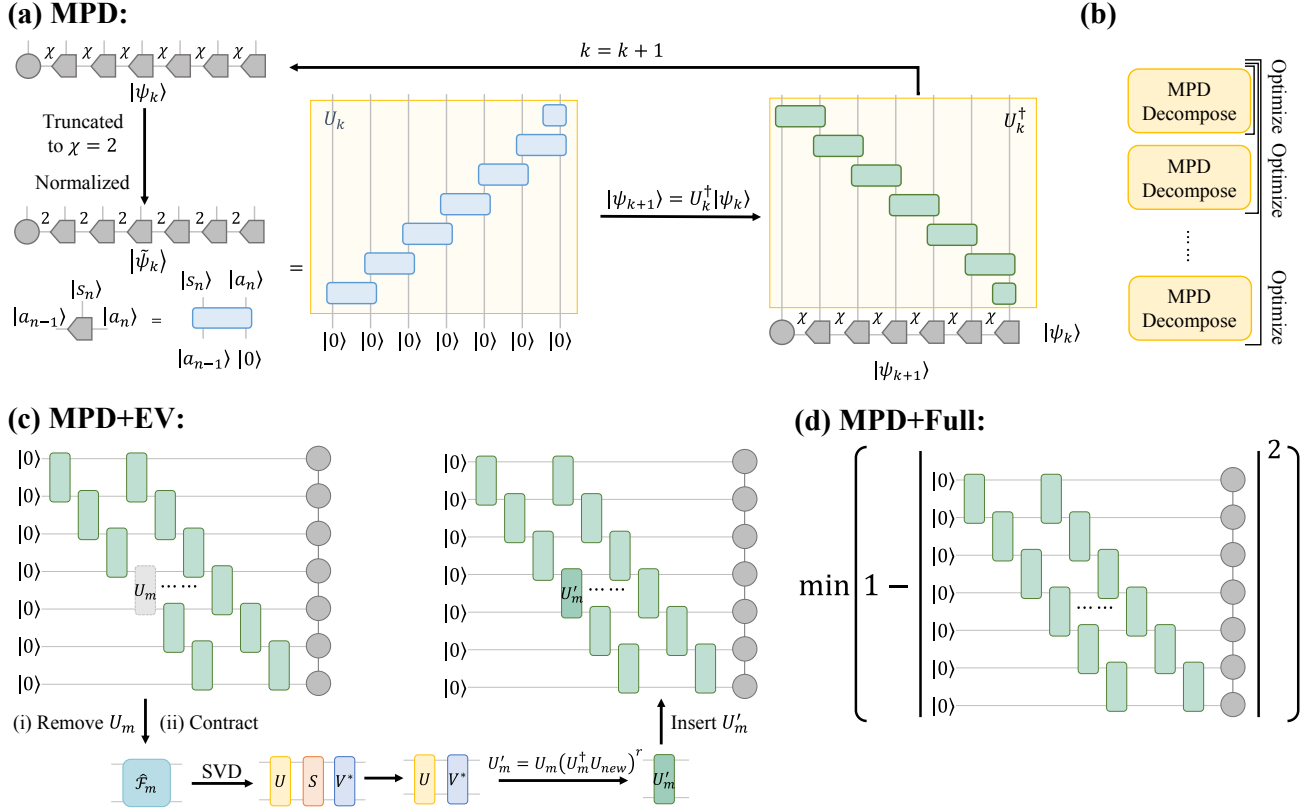}
    \caption{Schematic illustrations of approximate MPS decomposition methods based on multilayer staircase circuits. (a) MPD algorithm. At each layer, the residual MPS is truncated to bond dimension $\chi=2$ and exactly decomposed into a staircase layer of nearest-neighbor two-qubit gates. (b) Layerwise construction in the MPD+EV and MPD+Full algorithms. After a new MPD layer is appended, the accumulated circuit is optimized according to the corresponding algorithm. (c) Gatewise optimization in MPD+EV. The environment tensor $\hat{{\mathcal F}}_m$ of gate $\hat{U}_m$ is contracted, and its singular value decomposition determines the updated unitary $\hat{U}'_m$. (d) Full-circuit optimization in MPD+Full, where all variational gates are jointly optimized by minimizing the infidelity $1-F$.}
    \label{fig:protocols1}
\end{figure*}

We consider an $N$ sites quantum state represented in the open-boundary matrix product state (MPS) form,
\begin{equation}
\label{eq:mps}
\begin{aligned}
\ket{\Psi}=\sum_{\{s_n,a_n\}}A^{s_1}_{a_1}[1]A^{s_2}_{a_1,a_2}[2]\cdots A^{s_N}_{a_{N-1}}[N]\prod_{n=1}^{N}\ket{s_n} .
\end{aligned}
\end{equation}
Here, $s_n=1,\ldots,d$ labels the physical degree of freedom at site $n$, while $a_n=1,\ldots,\chi_n$ denotes the virtual index connecting the tensors at sites $n$ and $n+1$. The bond dimensions $\{\chi_n\}$ control the expressive capacity of the MPS and determine the maximum entanglement that can be encoded across the corresponding bipartitions. Compared with a full state-vector representation, which requires ${O}(d^N)$ coefficients, an MPS with maximum bond dimension $\chi_{\max}$ requires only ${O}(Nd{\chi^2_{\max}})$ parameters. Therefore, the MPS provides a compact representation for many-body states with moderate entanglement. 

\subsection{Exact mapping from MPS to quantum circuits}
\label{sec:exact_map}

A normalized MPS can be mapped exactly to a quantum circuit. Different choices of canonical form or orthogonality center lead to different compilation mappings and gate layouts, while leaving the represented quantum state unchanged. In this work, we use the right-canonical form.\cite{Schollwock2011}
Specifically, for the tensors $A^{s_n}_{a_{n-1},a_n}[n]$ with $n=2,3,\ldots,N-1$, the right-canonical condition is given by
\begin{equation}
\label{eq:right_canonical_bulk}
\sum_{s_n,a_n}A^{s_n}_{a_{n-1},a_n}[n]A^{s_n *}_{a_{n-1},a_n}[n]=\delta_{a_{n-1},a'_{n-1}} .
\end{equation}
For the right boundary tensor, this condition reduces to
\begin{equation}
\label{eq:right_canonical_boundary}
\sum_{s_N}A^{s_N}_{a_{N-1}}[N]A^{s_N*}_{a_{N-1}}[N]=\delta_{a_{N-1},a'_{N-1}} ,
\end{equation}
while the left boundary tensor satisfies the normalization condition
\begin{equation}
\label{eq:left_boundary_norm}
\sum_{s_1,a_1}A^{s_1}_{a_1}[1]A^{s_1*}_{a_1}[1]=1 .
\end{equation}
For each site, Eq.~\eqref{eq:right_canonical_bulk} exhibits that the tensor $A^{s_n}_{a_{n-1},a_n}[n]$ can be reshaped into an isometric matrix by combining the physical index $s_n$ and the right virtual index $a_n$ into a composite row index, namely $\mathcal{A}_{(s_n,a_n),a_{n-1}}[n] \equiv A^{s_n}_{a_{n-1},a_n}[n]$.

In the following, we consider qubit systems with local physical dimension $d=2$. To see the construction explicitly, consider first a local qubit tensor with dimensions $\chi\times 2\times\chi$, where $\chi=2^q$. In this case, $\mathcal{A}^{[n]}$ is a $2\chi\times\chi$ isometric matrix. Since the columns of $\mathcal{A}^{[n]}$ are orthonormal, $\mathcal{A}^{[n]}$ can be completed to a unitary matrix $U^{[n]}\in\mathbb{C}^{2\chi\times 2\chi}$ by adding $\chi$ additional orthonormal columns. This unitary completion is not unique, but it leaves the original isometry $\mathcal{A}^{[n]}$ unchanged. Consequently, the unitary matrix $U^{[n]}$ can be interpreted as an $(q+1)$-qubit gate,
see Figure~\ref{fig:circuit}(a).
For the boundary tensors, we introduce trivial virtual indices $a_0=a_N=1$ and write them as $A^{s_1}_{a_0,a_1}[1]$ and $A^{s_N}_{a_{N-1},a_N}[N]$. They can then be reshaped and completed to unitary gates in the same manner as the bulk tensors.

After each isometry $\mathcal A^{[n]}$ is completed to a unitary matrix $U^{[n]}$, the corresponding unitary gates are applied sequentially to the initial state $\ket{00\cdots0}$ according to the layout shown in Figure~\ref{fig:circuit}(a).
The $\ket{0}$ inputs select the $\mathcal A^{[n]}$ block of $U^{[n]}$.
Therefore, the sequential application of the unitary gates $U^{[n]}$ prepares the target MPS from the all-zero product state, and for an MPS with maximum bond dimension $\chi_{\max}$, the state can be represented by a circuit of $N$ multi-qubit gates, each acting on at most $\lceil \log_2 \chi_{\max}\rceil+1$ qubits. 

\subsection{Approximate compilation of MPS into quantum circuits}

In this section, we describe the five approximate MPS compilation methods investigated in this work, summarized in Table~\ref{tab:methods}.  The methods fall into two classes according to their circuit layouts. The first class comprises three methods that employ multiple staircase layers of two-qubit gates, as shown in Figure~\ref{fig:circuit}(b). Among these, the first is analytical and uses the exact decomposition of MPSs with $\chi_{\max}=2$~\cite{Ran2020}, whereas the other two optimize the local gates variationally. The second class are hybrid methods based on the layout shown in Figure~\ref{fig:circuit}(c). Each hybrid method begins with the same exact decomposition of the MPS into multi-qubit gates and then variationally approximates each multi-qubit gate by a sequence of local two-qubit gates according to different criteria.

\begin{table*}[t]
    \centering
    \begin{tabular}{cccccc}
    \hline\hline
        Algorithm & Layout & Classification & Cost function & Reference \\
        \hline
        MPD & Fig. \ref{fig:circuit}(b) & Analytical & NA &  Ref. \citenum{Ran2020} \\
        MPD+EV & Fig. \ref{fig:circuit}(b) & Variational & Fig. \ref{fig:protocols1}(c) &  Refs. \citenum{Rudolph2024,ShirakawaAQCE2024,EV2009} \\
        MPD+Full & Fig. \ref{fig:circuit}(b) & Variational & Fig. \ref{fig:protocols1}(d) &  Refs. \citenum{RanADQC2021,Schmelcher2026,GreenWang2025} \\
        DCD & Fig. \ref{fig:circuit}(c) & Hybrid & Fig. \ref{fig:protocols2}(a) & Ref. \citenum{YaoTingSu2025} \\
        DCE & Fig. \ref{fig:circuit}(c) & Hybrid & Fig. \ref{fig:protocols2}(b) &  this work \\
        \hline\hline
    \end{tabular}
    \caption{Summary of approximate compilation methods for compiling MPS into quantum circuits considered in this work. The MPD+Full algorithm implemented here is not identical to the algorithms in Refs. \citenum{RanADQC2021,Schmelcher2026,GreenWang2025}; however, they share the common feature that full optimization is performed to obtain the optimal local unitary gates.
    }
    \label{tab:methods}
\end{table*}

\subsubsection{Analytical matrix product disentanglers (MPD) algorithm}

The matrix product disentanglers (MPD) algorithm\cite{Ran2020}
is illustrated in Figure~\ref{fig:protocols1}(a). The iterative disentangling procedure is initialized with the target MPS, $\ket{\psi_1} \equiv \ket{\psi_\mathrm{target}}$. At iteration $k$, the current MPS $\ket{\psi_k}$ is first truncated to bond dimension $\chi=2$ and then normalized, yielding the truncated MPS $|\tilde{\psi}_k\rangle$. As discussed in Sec.~\ref{sec:exact_map}, the normalized truncated MPS $\lvert\tilde{\psi}_k\rangle$, with bond dimension $\chi=2$, can be mapped exactly to a single staircase layer, $U_k=\prod_{i=1}^{N-1}U_{k,i}$, of $N-1$ nearest-neighbor two-qubit gates, where $U_{k,i}$ is a $4\times4$ unitary acting on qubits $i$ and $i+1$. The rightmost single-qubit gates arising in the exact mapping are absorbed into adjacent two-qubit gates. Applying the adjoint layer $U_k^\dagger$ to the MPS $\ket{\psi_k}$ gives the next MPS, $\ket{\psi_{k+1}}=U_k^\dagger \ket{\psi_k}$. Repeating this procedure for $L$ iterations gives a sequence of disentangling layers $\{U_1^\dagger,U_2^\dagger,\ldots,U_L^\dagger\}$. Thus, the target MPS is then approximately prepared by $U_1U_2\cdots U_L\ket{00\cdots0}$.
We classify the MPD algorithm as analytical, since no optimization is involved.

In our implementation, we set the maximum number of iterations in the MPD algorithm to $L_{\max}=25$. After applying each inverse layer $U_k^\dagger$, we truncate the resulting MPS to a maximum bond dimension of $D=4\chi_{\max}$, where $\chi_{\max}$ denotes the maximum bond dimension of the target MPS $\ket{\psi_{\mathrm{target}}}$. This cutoff is necessary, because the present work considers relatively large number of qubits $N$ and layers $L_{\max}$. Without this restriction, the bond dimensions of the intermediate MPS may grow rapidly during the iterative application of the disentangling layers, making the computation prohibitively expensive.
The detailed algorithm is summarized in Algorithm 1 in Supporting Information.

\begin{figure*}[t]
    \centering
    \includegraphics[width=0.98\textwidth]{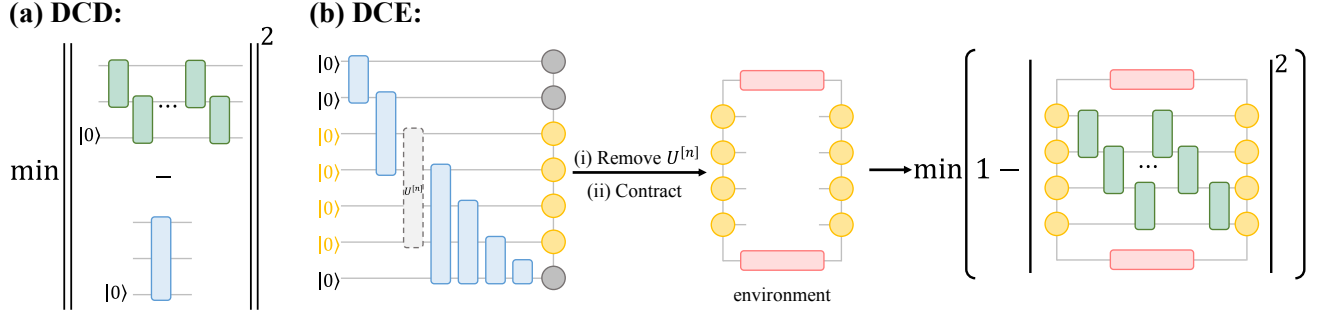}
    \caption{Schematic illustrations of two divide-and-conquer methods for approximate MPS decomposition based on multi-qubit gate decomposition. The multi-qubit gates obtained from the exact MPS decomposition in Figure~1(a) are independently decomposed into two-qubit circuits as shown in Figure~1(c), allowing the decompositions to be performed in parallel. (a) In the DCD algorithm, each multi-qubit gate is approximated by minimizing the squared Frobenius distance between the target gate and the corresponding two-qubit circuit. (b) In the DCE algorithm, each multi-qubit gate is approximated by minimizing the state infidelity after replacing only the target gate with its two-qubit circuit while keeping all other multi-qubit gates exact. The local objective is evaluated using the corresponding environment tensor network.}

    \label{fig:protocols2}
\end{figure*}

\subsubsection{Layer-by-layer variational optimization methods}
The MPD algorithm relies only on the exact decomposition of MPS with bond dimension $\chi=2$. 
An alternative strategy is to construct a multi-layer circuit of two-qubit gates through variational optimization, as illustrated in Figure~\ref{fig:protocols1}(b). In this strategy, the gates are first initialized using a simpler algorithm, typically the MPD algorithm. The optimization is then performed in a layer-by-layer manner. First, the initial layer of two-qubit gates is generated by the reference algorithm and optimized until convergence or until the maximum number of iterations is reached. A new layer, again initialized by the reference algorithm, is subsequently appended to the circuit, after which all existing layers are optimized together. Repeating this procedure yields a multi-layer circuit composed of two-qubit gates. 

{\bf MPD+EV optimization algorithm}
A widely used and effective algorithm optimizes the circuit through sequential single-gate updates\cite{Rudolph2024,ShirakawaAQCE2024}, following the tensor-network optimization scheme introduced by Evenbly and Vidal~\cite{EV2009}. In this work, we adopt the $\mathrm{Iter}[D_iO_{\mathrm{all}}]$ protocol described in Ref.~[\citenumtext{Rudolph2024}], with only minor implementation-level modifications. The gates are initialized using the MPD  algorithm. In each local step, one two-qubit gate $U_m$ is optimized with all other gates kept fixed.
The objective is to maximize the fidelity
\begin{equation}
F=
\left|\bra{\psi_{\mathrm{target}}}U\ket{00\cdots0}\right|^2 .
\end{equation}
To update a single two-qubit gate $U_m$, we first construct its environment tensor by
contracting all tensors except $U_m$, as illustrated in 
Figure~\ref{fig:protocols1}(c):
\begin{equation}
\hat{\mathcal F}_m=\operatorname{Tr}_{\bar{U}_m}\left[
\prod_{i=M}^{m+1} U_i^\dagger\ket{\psi_{\mathrm{target}}}
\bra{00\cdots0}\prod_{j=1}^{m-1} U_j^\dagger\right],
\label{eq:environment_tensor}
\end{equation}
where $\operatorname{Tr}_{\bar U_m}$ denotes the contraction over all indices except those connected to $U_m$. The tensor $\hat{\mathcal F}_m$ gives the local environment, or equivalently the local gradient of the overlap with respect to $U_m$~\cite{Torre2024}. Since $\hat{\mathcal F}_m$ is generally not unitary, the updated gate is chosen as the closest unitary matrix obtained from the singular-value decomposition of $\hat{\mathcal F}_m=\mathcal U\mathcal S\mathcal V^\dagger$, such that $U_{\mathrm{new}}=\mathcal U \mathcal V^\dagger$. We then update the gate by\cite{Rudolph2024}
\begin{equation}
U_m^{'}=U_m\left(U_m^\dagger U_{\mathrm{new}}\right)^r,
\label{eq:polar_projection}
\end{equation}
with the learning rate $r=0.6$, as shown in Figure~\ref{fig:protocols1}(c).

In our implementation, we set the maximum number of layers $L_{\max}=25$ and use a maximum bond dimension $D=4\chi_{\max}$ for obtaining the environments during the optimization. We perform full forward sweeps to optimize the local gates.
We monitor the infidelity after every 20 sweeps, with an additional evaluation after the first sweep to initialize the convergence reference. Optimization at the current depth stops when the relative change between two successive monitored infidelities falls below $10^{-3}$. After this optimization stage, no further layer is added if the final monitored infidelity satisfies $1-F<10^{-7}$. For details, see Algorithm 2 in Supporting Information.

{\bf MPD+Full optimization algorithm}
For the MPD+Full algorithm, each layer is initialized using MPD. 
The first layer is optimized using MPD+EV, whereas from the second layer onward, all variational parameters are jointly optimized by minimizing the infidelity $1-F$ using the Adam optimizer, as illustrated in Figure~\ref{fig:protocols1}(d). The unitary constraint is enforced using a QR decomposition during optimization. In each optimization step, the infidelity is evaluated by directly contracting the full tensor network corresponding to the overlap $\bra{\psi_\mathrm{target}}U\ket{00\cdots 0}$, without imposing any upper bound on the intermediate bond dimensions. For this full-network contraction, we employ the \textsc{Quimb} package\cite{gray2018quimb} and use the automatic high-quality (\texttt{auto-hq}) contraction-path strategy. 
The implemented MPD+Full algorithm is summarized in Algorithm 3 in Supporting Information.

Since both the automatic search for an efficient contraction path and the subsequent exact contraction of the full tensor network become increasingly expensive as the number of layers increases, only relatively shallow circuits are considered for MPD+Full in this work. Specifically, we restrict the MPD+Full calculations to $L_{\max}=10$, which is smaller than that used in the other methods, due to the cost of evaluating the infidelity.

The MPD, MPD+EV, and MPD+Full algorithms are evaluated by computing the final infidelities of the generated circuits using an MPS representation with a cutoff for the bond dimension $\chi_{\mathrm{cut}}=8\chi_{\max}$. 
This larger bond dimension is used to reduce the truncation error in the final infidelity evaluation. The convergence of the MPD+EV infidelity with respect to the bond dimension used in the final evaluation is examined in 
Figure S1 in Supporting Information, where the generated circuit is kept unchanged. The results show that the infidelity evaluated with $\chi_{\mathrm{cut}}=8\chi_{\max}$ has already converged compared with those obtained using larger cutoffs.

\subsubsection{Divide-and-conquer methods}
For the hybrid methods considered in this work, each multi-qubit gate is approximated independently by local two-qubit gates, allowing all decompositions to be performed in parallel. We refer to such methods as divide-and-conquer (DC) methods.
Since the DC methods require the virtual dimensions of the MPS tensors to be powers of two, we pad each input MPS accordingly. Each tensor $A^{[n]}$ of shape $(\chi_{n-1},2,\chi_n)$ is first embedded into a zero-padded tensor $\bar{A}^{[n]}$ of shape $(2^{q_{n-1}},2,2^{q_n})$, where $q_n=\lceil\log_2\chi_n\rceil$, by retaining the original entries in the corresponding block and setting all additional entries to zero. The zero-padded MPS is then re-canonicalized by first left-canonicalizing and subsequently right-canonicalizing. This procedure leaves the represented physical state unchanged and restores the canonical form of the padded MPS. The resulting canonical tensors denoted by $\widetilde{A}[n]$ are used in the subsequent construction of the multi-qubit gates.
After a gauge transformation, each tensor can be completed to a multi-qubit gate with determinant $+1$. For the real-valued MPSs considered in the present work, this completion can be chosen to be real orthogonal. Together, these properties ensure compatibility with the use of $\mathrm{SO}(4)$ two-qubit gate described below. 

{\bf DC distance (DCD) algorithm} The algorithm used in Ref.~[\citenumtext{YaoTingSu2025}] variationally minimizes the squared Frobenius distance between the target multi-qubit gate and the corresponding circuit of two-qubit gates, evaluated on the input subspace selected by the $\ket{0}$ inputs, see Figure~\ref{fig:protocols2}(a). 
Each optimization run uses the limited-memory BFGS (L-BFGS) algorithm with a learning rate of $0.1$ and a line search satisfying the strong Wolfe conditions. For each mapped multi-qubit gate, we consider the maximum number of circuit layers to $L_{\max}=25$, stopping earlier if the minimum loss obtained at the current depth falls below $10^{-7}$. Details of the DCD algorithm are summarized in Algorithm 4 in Supporting Information.

{\bf DC environment (DCE) algorithm} The DCD algorithm approximates each multi-qubit unitary gate without using information from the full target state $|\psi_{\mathrm{target}}\rangle$.
To incorporate this information, we propose a divide-and-conquer algorithm that uses the environment constructed from $|\psi_{\mathrm{target}}\rangle$, referred to as the DCE algorithm, see Figure \ref{fig:protocols2}(b).
 In this protocol, each multi-qubit gate $U^{[n]}$ is approximated by a circuit of local two-qubit gates. We optimize each circuit using a layer-by-layer strategy~\cite{2024Xiao} and set the maximum number of layers to $L_{\max}=25$. The two-qubit gates
$U^{[n]}_1,U^{[n]}_2,\ldots,U^{[n]}_i$ within this block are contracted and regarded as an effective $(q_n+1)$-qubit gate $\widetilde U^{[n]}$. The approximation is optimized by minimizing the infidelity between the target state and the state obtained after replacing only $U^{[n]}$ by ${\widetilde U}^{[n]}$, while all other multi-qubit gates are kept exact. 
For compact notation, we define
$\bra{L_n}=\bra{\psi_{\mathrm{tar}}}U^{[N]}U^{[N-1]}\cdots U^{[n+1]}$ and $\ket{R_n}=U^{[n-1]}\cdots U^{[1]}\ket{00\cdots0}$. The local loss function is then written as
\begin{equation}
\mathcal L_n=1-\lvert\bra{L_n}\widetilde U^{[n]}\ket{R_n}\rvert^2,
\label{eq:DCE_loss}
\end{equation}
as illustrated in the right panel of Figure~\ref{fig:protocols2}(b). To construct the environment for ${\widetilde U}^{[n]}$, we first contract the exact gates before and after the target block into two effective MPSs. We then contract all sites except those acted on by ${\widetilde U}^{[n]}$, yielding the local environment tensor network shown in the middle panel of Figure~\ref{fig:protocols2}(b). The contraction of the first $n-1$ sites gives the front environment, while the contraction of last $N-n-q_n$ sites gives the back environment.
The two-qubit gates in ${\widetilde U}^{[n]}$ are then optimized by contracting this local tensor network and minimizing $\mathcal L_n$. In this construction, the largest bond dimension is bounded by $\chi_{\max}$, so the loss function can be evaluated without truncation.
In the Appendix, we analyze the relationship between
local and global infidelities, and show that in order to achieve a global error $\epsilon_{\mathrm{global}}$,
the local error does not need to decrease exponentially with respect to the system size.

The DCE algorithm optimizes the two-qubit gates on the special orthogonal manifold. Each local gate is restricted to $U^{[n]}_i\in \mathrm{SO}(4)$, and the conjugate-gradient algorithm is used for the manifold optimization.\cite{2016pymanopt} The manifold optimization with the conjugate-gradient optimizer also achieves comparable results to the exponential parameterization with the L-BFGS optimizer used in the DCD algorithm, see Figure S2 in Supporting Information. Although the exponential parameterization with the L-BFGS optimizer can also be applied to DCE and generally gives similar results, we find that in general the manifold optimization with the conjugate-gradient optimizer provides more robust performance. 
For details of the DCE algorithm, see Algorithm 5 in Supporting Information.

\section{Results}\label{sec:results}
\subsection{Benchmark of different MPS compilation methods}

\begin{figure*}[t]
    \centering
    \includegraphics[width=0.98\textwidth]{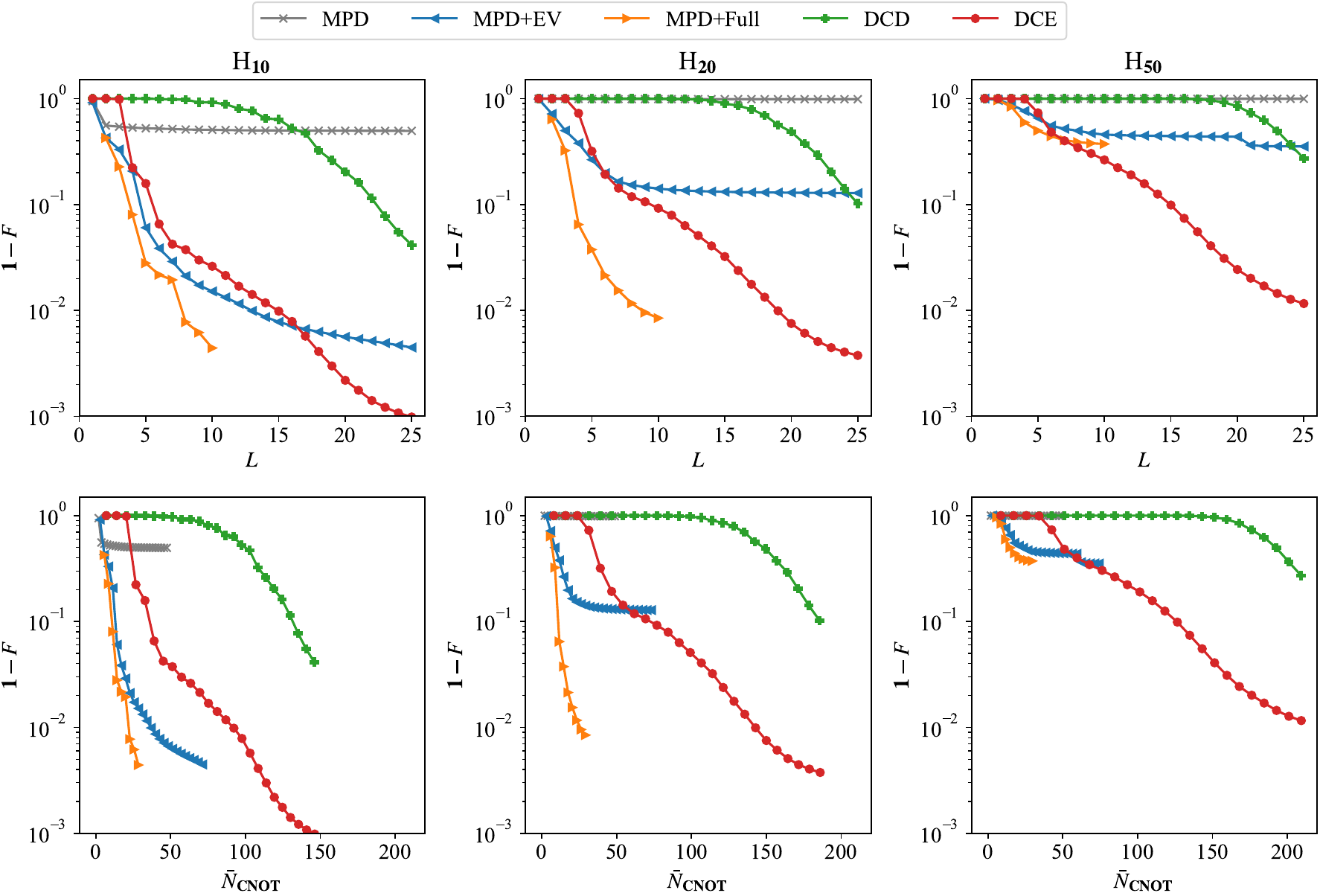}
    \caption{Benchmark comparison of representative MPS decomposition methods for the $\mathrm{H}_{10}$, $\mathrm{H}_{20}$, and $\mathrm{H}_{50}$ chains, corresponding to $N=20$, $40$, and $100$ qubits, respectively, at an interatomic distance of $1.0~\text{\AA}$ and a fixed bond dimension of $\chi_{\max}=32$. The upper panels show the infidelity as a function of the number of layers $L$, while the lower panels show the same results as a function of the average number of CNOT gates per qubit, $\bar N_\mathrm{CNOT}=N_{\mathrm{CNOT}}/N$.
    }
    \label{fig:Hchain_dcut32}
\end{figure*}

\begin{figure*}[t]
    \centering
    \includegraphics[width=0.98\textwidth]{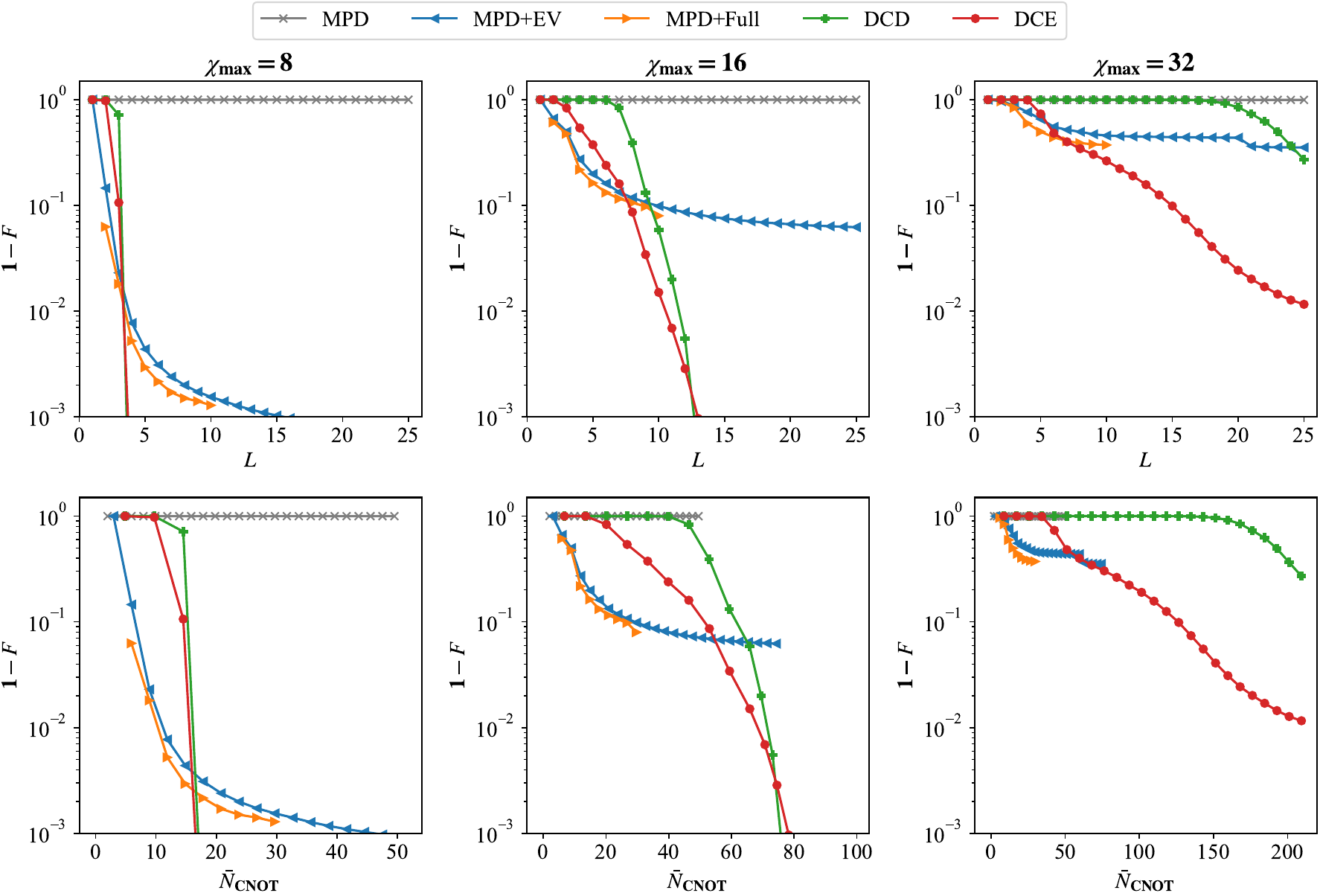}
    \caption{Benchmark comparison of representative MPS decomposition methods for the $\mathrm{H}_{50}$ chain at an interatomic distance of $1.0~\text{\AA}$ with bond dimensions $\chi_{\max}=8$, $16$, and $32$. The upper panels show the infidelity as a function of the number of layers $L$, while the lower panels show the same results as a function of the average number of CNOT gates per qubit, $\bar N_\mathrm{CNOT}=N_{\mathrm{CNOT}}/N$. 
    }
    \label{fig:Hchain_H50}
\end{figure*}

\begin{figure*}[t]
    \centering
    \includegraphics[width=0.98\textwidth]{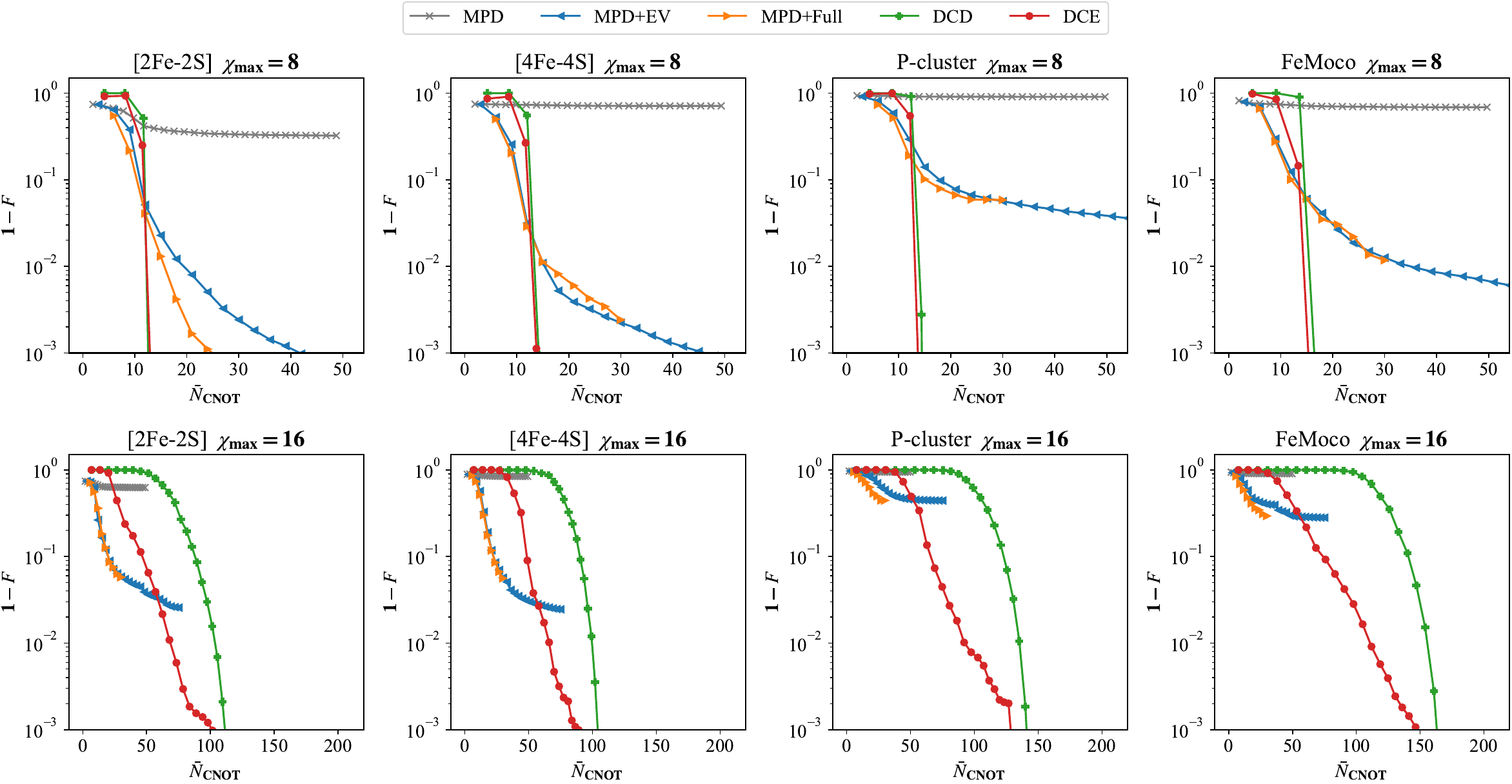}
    \caption{Benchmark comparison of representative MPS decomposition methods for strongly correlated iron-sulfur systems, including [2Fe-2S], [4Fe-4S], P-Cluster, and FeMoco. The upper and lower panels correspond to MPSs with bond dimensions $\chi_{\max}=8$ and $16$, respectively. The infidelity is shown as a function of the average number of CNOT gates per qubit, $\bar N_\mathrm{CNOT}=N_{\mathrm{CNOT}}/N$. 
    }
    \label{fig:iron_infi}
\end{figure*}

\begin{figure*}[t]
    \centering
    \includegraphics[width=0.98\textwidth]{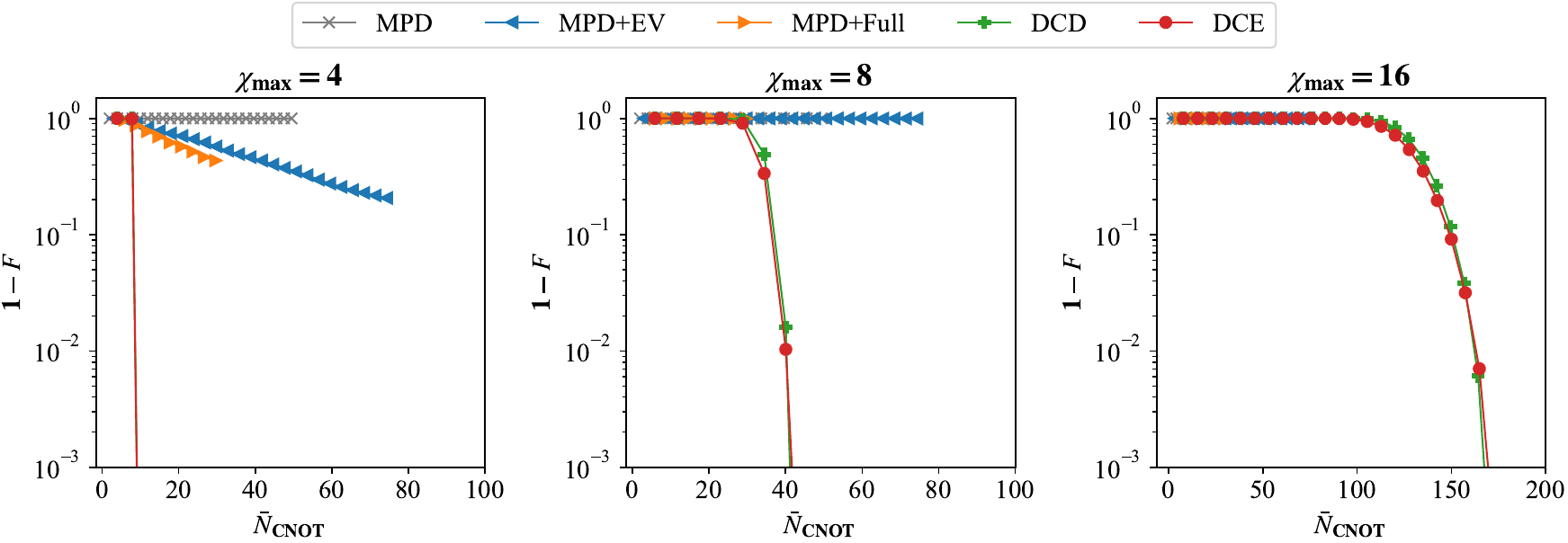}
    \caption{Benchmark comparison of representative MPS decomposition methods for random MPSs with $N=100$ and bond dimensions $\chi_{\max}=4$, $8$, and $16$. The infidelity is shown as a function of the average number of CNOT gates per qubit, $\bar N_\mathrm{CNOT}=N_{\mathrm{CNOT}}/N$. 
    }
    \label{fig:random_infi}
\end{figure*}

\begin{figure}[t]
    \centering
    \includegraphics[width=0.48\textwidth]{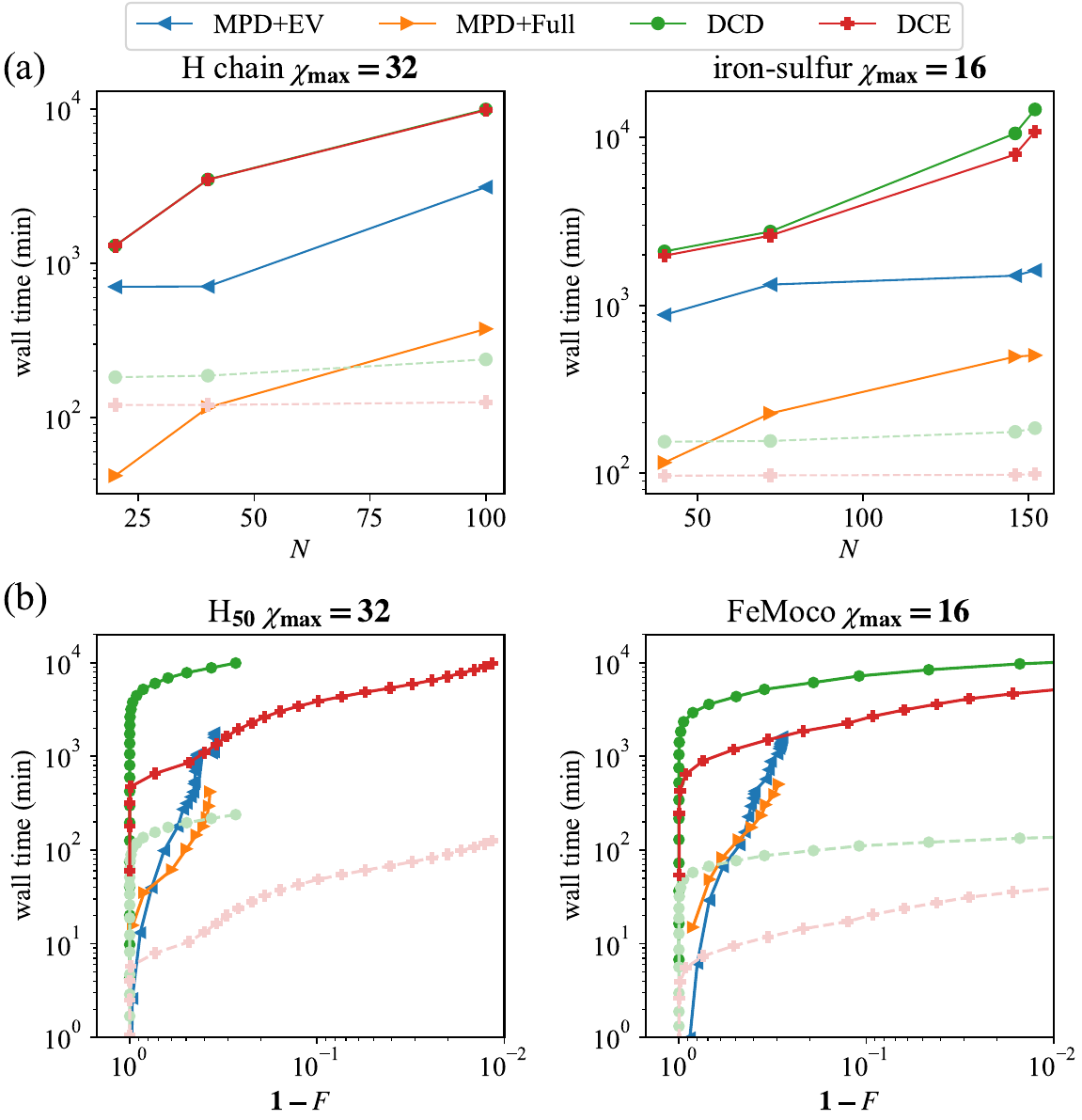}
    \caption{Comparison of computational performance for representative MPS decomposition methods.
    (a) Wall time as a function of the number of qubits $N$ for hydrogen chains with $\chi_{\max}=32$ and iron-sulfur systems with $\chi_{\max}=16$ with $L_{\max}=10$
    for MPD+Full and $L_{\max}=25$ for the other methods. 
    (b) Wall time as a function of the target infidelity $1-F$ for the $\mathrm{H}_{50}$ chain at $1.0~\text{\AA}$ with $\chi_{\max}=32$ and for FeMoco with $\chi_{\max}=16$.
    Solid lines denote the measured wall time, obtained using 8 CPU cores for MPD-based methods and 6 CPU cores for DC methods. Dashed lines indicate the ideal wall time assuming perfect parallelization of all independent tasks.}
    \label{fig:properties}
\end{figure}

\subsubsection{MPS for hydrogen chains}
We benchmark different MPS compilation methods for linear hydrogen chains in the STO-3G basis using orthogonal atomic orbitals (OAOs), including \ce{H10}, \ce{H20}, and \ce{H50}. The MPSs are first generated by DMRG in the spatial-orbital ($d=4$) representation using the \textsc{FOCUS} package\cite{li2021expressibility,EMO},
and then converted into spin-orbital representation ($d=2$)
with bond dimension $\chi_{\max}$.
Figure~\ref{fig:Hchain_dcut32} summarizes the results
for \ce{H10}, \ce{H20}, and \ce{H50} with $\chi_{\max}=32$,
while Figure~\ref{fig:Hchain_H50} compares the results
for \ce{H50} with $\chi_{\max}=8$, $16$, and $32$. 
In Figures~\ref{fig:Hchain_dcut32} and \ref{fig:Hchain_H50}, the top panel represents the infidelity as a function of the number of layers $L$, and the bottom panel exhibits the same infidelity as a function of the average number of CNOT gates per site, $\bar N_\mathrm{CNOT}=N_{\mathrm{CNOT}}/N$, where $N_{\mathrm{CNOT}}$ is the total number of CNOT gates. To enable a consistent resource comparison across the different methods, we convert the numbers of \(\mathrm{U}(4)\) and \(\mathrm{SO}(4)\) gates into CNOT counts using their maximum exact decomposition costs. Specifically, each generic \(\mathrm{U}(4)\) gate is counted as three CNOT gates, whereas each generic \(\mathrm{SO}(4)\) gate is counted as two CNOT gates, assuming arbitrary single-qubit rotations are available.

As shown in Figure~\ref{fig:Hchain_dcut32} for various system sizes $N$ at a fixed bond dimension of $\chi_{\max}=32$, MPD provides only limited improvement with additional layers, and its infidelity does not fall below $0.5$ for any of the systems considered.
For \ce{H10}, MPD+EV and MPD+Full are efficient, reducing the infidelity below $10^{-2}$ with $\bar N_\mathrm{CNOT}<30$ and requiring substantially smaller $\bar N_\mathrm{CNOT}$ values than the other three methods. 
As the system size increases, however, the behavior of MPD+EV and MPD+Full algorithms becomes more differentiated. For \ce{H20}, MPD+Full still reaches infidelities below $10^{-2}$ with a small number of gates, whereas MPD+EV already exhibits a clear saturation at a higher infidelity. For the larger \ce{H50}, both MPD+EV and MPD+Full become less effective, with their infidelities remaining at relatively large values. These results indicate that the performance of MPD+EV and MPD+Full algorithms is strongly affected by system size.

DCD also fails to reduce the infidelity below $10^{-2}$ within $L_{\max}=25$ layers in all cases. Nevertheless, its behavior is different from that of MPD+EV and MPD+Full. The infidelity starts to decrease only at sufficiently large $L$, and unlike the behavior observed for the MPD-based methods, no clear plateau is observed at larger $L$. The DCE algorithm exhibits the most consistent good performance across different system sizes. For all systems considered, DCE reduces the infidelity to values close to or below $10^{-2}$ within $L_{\max}=25$ layers.
It also consistently outperforms DCD, yielding lower infidelities at the same $\bar N_\mathrm{CNOT}$ over the full range of system sizes. 
Compared with MPD+EV and MPD+Full, DCE is not necessarily the most efficient algorithm in terms of $\bar N_\mathrm{CNOT}$ for small systems with $O(10)$ qubits. However, it clearly becomes more advantageous for larger systems with $O(10^2)$ qubits.

We observe similar behaviors of different methods 
for \ce{H50} with various bond dimensions in Figure~\ref{fig:Hchain_H50}. 
MPD still gives bad infidelities close to unity across all bond dimensions considered. 
At small $\bar N_\mathrm{CNOT}$, MPD+EV and MPD+Full give relatively low infidelities. DCE is generally superior to DCD at a given $\bar{N}_{\mathrm{CNOT}}$ and achieves low infidelity for all systems considered within $L_{\max}=25$ layers.
The advantages of DCE over the other methods become more apparent for the larger bond dimensions considered here.

\subsubsection{MPS for iron-sulfur clusters}
We further benchmark different MPS compilation methods on MPS for iron-sulfur clusters, which represent chemically more realistic systems with strong electronic correlation. Specifically, we consider [2Fe(III,III)-2S]~\cite{2Fe-2S}, [4Fe(III,III,II,II)-4S]~\cite{2Fe-2S}, P-cluster~\cite{P-Cluster}, and FeMoco~\cite{FeMoco}, following the molecular structures and active-space definitions reported in Ref.~\cite{EMO}. The corresponding active spaces are $(30e,20o)$, $(54e,36o)$, $(114e,73o)$, and $(113e,76o)$, respectively. For the present benchmarks, the $d=2$ MPSs with $\chi_{\max}=8$ and $16$ were obtained from the corresponding $d=4$ MPSs 
with maximum bond dimensions of $4$ and $9$, respectively, 
compressed from large bond dimension DMRG calculations
in the spin-restricted EMO basis~\cite{EMO},
followed by conversion to $d=2$ with further truncation to target bond dimensions.

The results obtained with different methods are shown in Figure \ref{fig:iron_infi}. In all cases, the MPD algorithm fails to reduce the infidelity.
For $\chi_{\max}=8$, both DCE and DCD achieve very low infidelities already at $\bar N_\mathrm{CNOT}=20$, consistent with the preceding benchmark results. For MPD+EV and MPD+Full, the infidelity can be reduced to approximately $0.01$ for most systems, except for the P-cluster system, where noticeably larger values remain. 
For $\chi_{\max}=16$, the differences among the methods become more evident. MPD+EV and MPD+Full give relatively large infidelities, especially for P-cluster and FeMoco, where the minimum infidelities for both algorithms remain around $0.4$ and $0.2$, respectively. For the DCD and DCE algorithms, the infidelity can be reduced to relatively low values for all systems. At fixed $\bar N_\mathrm{CNOT}$, DCE consistently gives lower infidelities than DCD.

\subsubsection{Random MPS}
To provide a less problem-specific test of the behaviors of different compilation algorithms, we further examine random real-valued MPS benchmark states with $N=100$ qubits and bond dimensions $\chi_{\max}$=4, 8, and 16, see Figure~\ref{fig:random_infi}. The target states are generated from the all-zero product state by sequentially applying random orthogonal multi-qubit gates, following the circuit structure in Figure~\ref{fig:circuit}(a). In the bulk of the chain, each gate acts on $\log_2\chi_{\max}+1$ qubits, while smaller gates are used near the boundaries to match the available virtual dimensions. Each gate is independently sampled from the Haar orthogonal ensemble. All numerical settings are the same as before.

The MPD algorithm remains ineffective across all three bond dimensions, with infidelities close to unity throughout. The MPD+Full and MPD+EV algorithms give only limited improvement for $\chi_{\max}=4$, with final infidelities still above $0.1$. Their performance deteriorates rapidly for $\chi_{\max}=8$ and $\chi_{\max}=16$, where the infidelity remains close to unity. In contrast, DCD and DCE show nearly identical performance over the range considered, with both algorithms reaching very low infidelities at sufficiently large $L$ or $\bar N_\mathrm{CNOT}$. For example, for $\chi_{\max}=4$, both algorithms reach infidelities below $10^{-5}$ at $L=3$. At this layer number, the local decomposition contains $2\times 6L=36$ real parameters, which is larger than the $2^{2\times3-1}-\frac{2^{3-1}(2^{3-1}+1)}{2}=22$ real parameters required to exactly describe the relevant three-qubit block selected by a one-qubit $\ket{0}$ input. Considering the parameter count, the ansatz may already contain enough parameters to represent the relevant block of a three-qubit gate, consistent with the low infidelities observed numerically.

Overall, for all systems considered in this work, DCE consistently yields low infidelities with a moderate number of layers. This indicates that DCE is more robust across different systems and bond dimensions considered here.

\subsection{Timing and scalability}
Apart from the final infidelity and gate count, we next compare the wall time with respect to system size $N$ and achieved infidelity $1-F$. These properties are important in practical applications, where a useful method needs to reach the desired accuracy within a feasible wall time.
MPD+EV relies on sequential local gate updates and scales approximately as $\mathcal{O}(N 
N_{\mathrm{iter}} L_{\max}^2 \chi_{\max}^3 )$. MPD+Full uses a global objective whose tensor contraction cost grows rapidly with $L$. These MPD-based methods therefore have limited task level parallelism, with parallel speedups mainly coming from linear algebra operations such as SVD and tensor contractions.
In contrast, the DC methods split the decomposition into $N$ independent local tasks, and both these tasks and the six independent optimization runs for each task can be executed in parallel. Under the ideal parallel execution, the wall time is ultimately determined by the slowest local decomposition, which scales as $\mathcal{O}[N_{\mathrm{iter}}L_{\max}(\chi_{\max}^3+L_{\max}\chi_{\max}^2\log_2\chi_{\max})]$, without an explicit dependence on $N$. 

Figure~\ref{fig:properties}(a) shows the wall-clock time as a function of system size for the hydrogen-chain benchmarks with $\chi_{\max}=32$ and the iron-sulfur cluster benchmarks with $\chi_{\max}=16$. Here, the layer number is fixed at $L_{\max}=25$ for all methods except MPD+Full, for which $L_{\max}=10$ is used. 
All timing benchmarks were performed on a compute node equipped with two Intel Xeon Gold 6248R CPUs (3.00 GHz), providing 48 physical CPU cores and 503 GB RAM. For MPD+EV and MPD+Full, we report the runtime of a single calculation performed with 8 CPU cores. For DCD and DCE, we report the
wall time of all tasks and the maximum wall time among all local decompositions of the multi-qubit gates. Each local decomposition uses 6 CPU cores, with one core assigned to each independent optimization run within a task
for decomposing a multi-qubit gate. 
It is observed that the maximum wall time of the individual
tasks is almost independent of $N$, as it mainly depends
on $\chi_{\max}$. While the total wall time of DC methods is longer than MPD+EV, they can be made more efficient with parallelization.

While Figure~\ref{fig:properties}(a) compares the wall time
of different methods for fixed $L_{\max}$, a more practical measure for scalability is the wall time for achieving a fixed infidelity $1-F$, since different methods have very different convergence behaviors
for infidelity as we have demonstrated in the previous subsection. Figure~\ref{fig:properties}(b) summarizes the wall-clock time as a function of the achieved infidelity for $\mathrm{H}_{50}$ with $\chi_{\max}=32$ and FeMoco with $\chi_{\max}=16$. Both MPD+EV and MPD+Full become increasingly expensive at lower infidelities. For these systems, they cannot achieve an infidelity below 0.1
with the present $L_{\max}$. In contrast,
the increase in wall time remains relatively slower for DCD and DCE, as lower infidelities are targeted.
Consequently, for larger systems, the DC methods achieve low infidelities more efficiently than MPD+EV and MPD+Full.

\section{Conclusion}\label{sec:conclusion}
We have systematically compared analytical, variational, and hybrid MPS compilation methods using hydrogen chains and iron-sulfur clusters including the P-cluster and FeMoco. The benchmark considers state infidelity, CNOT gate count, classical runtime, and parallel scalability over different system sizes and MPS bond dimensions. 
The results provide concrete guidance for practical method selection. MPD followed by  variational refinement, such as MPD+EV or MPD+Full, can often achieve the target accuracy with fewer CNOT gates for some commonly encountered molecular systems with both $O(10)$ qubits and $O(10)$ bond dimensions, or when an infidelity of order $10^{-1}$ is acceptable. 
For larger systems with $O(10^2)$ qubits and relatively higher bond dimensions $O(10\text{-}10^2)$, direct variational optimization often becomes increasingly expensive, and the improvement in fidelity can become progressively slower. In this regime, the introduced DCE compilation is more reliable, particularly when circuit infidelities around or below $10^{-2}$ are required.
Based on these observations, a practical workflow is to first test MPD with variational refinement under a limited gate budget and switch to DCE if the required fidelity is not reached.

For iron–sulfur clusters, our results demonstrate that combining EMOs\cite{EMO} with DCE provides a resource-efficient route to initial-state preparation. EMOs enable compact MPS representations of approximate ground states at modest bond dimensions, while DCE permits their efficient parallel compilation into quantum circuits. Future developments should incorporate symmetry preservation, adaptive circuit architectures, hardware-native connectivity, and error-aware resource allocation to enable end-to-end preparation of strongly correlated molecular states for quantum simulation. When integrated with QPE, these techniques could form a practical toolkit for accurately treating strongly correlated molecules and generating high-quality reference data in regimes where classical methods alone become unreliable.

\appendix
\setcounter{equation}{0}
\renewcommand{\theequation}{A\arabic{equation}}

\section*{Appendix: Relationship between Local and Global Infidelities}
This section derives a conservative bound relating the local infidelities to the infidelity of the assembled circuit. Consider the target state and the approximate state obtained by replacing each unitary $U^{[n]}$ with its approximation $\widetilde U^{[n]}$,
\begin{equation}
\begin{aligned}
\lvert\psi_{\mathrm{target}}\rangle
&=
U^{[N]}U^{[N-1]}\cdots U^{[1]}
\lvert00\cdots0\rangle ,
\\
\lvert\widetilde{\psi}\rangle
&=
\widetilde U^{[N]}\widetilde U^{[N-1]}
\cdots\widetilde U^{[1]}
\lvert00\cdots0\rangle .
\end{aligned}
\end{equation}
The global fidelity and infidelity are defined as
\begin{equation}
F=\left|\langle\psi_{\mathrm{target}}\vert\widetilde{\psi}\rangle\right|^2,\qquad \mathcal{L}=1-F.
\end{equation}
For the $n$-th unitary, the corresponding local fidelity and infidelity are
\begin{equation}
F_n=1-\mathcal{L}_n=\left|\langle L_n\rvert\widetilde U^{[n]}\lvert R_n\rangle\right|^2.
\label{eq:12}
\end{equation}
Here, $\lvert L_n\rangle$ and $\lvert R_n\rangle$ are the left and right states defined in the main text. To relate the local infidelities to the global infidelity, we use the Fubini--Study distance. For any two normalized pure states, it is defined as
\begin{equation}
d_{\mathrm{FS}}\!\left(\lvert a\rangle,\lvert b\rangle\right)=\arccos\left|\langle a\vert b\rangle\right|.
\label{eq:fs-distance}
\end{equation}
The Fubini--Study distance defines a metric on the projective Hilbert space and therefore satisfies the triangle inequality,
\begin{equation}
d_{\mathrm{FS}}\!\left(\lvert a\rangle,\lvert c\rangle\right)
\leq d_{\mathrm{FS}}\!\left(\lvert a\rangle,\lvert b\rangle\right)
+d_{\mathrm{FS}}\!\left(\lvert b\rangle,\lvert c\rangle\right).
\label{eq:triangle}
\end{equation}

We introduce a sequence of intermediate states,
\begin{equation}
\lvert\Phi_n\rangle=\widetilde U^{[N]}\cdots\widetilde U^{[n+1]}
U^{[n]}\cdots U^{[1]}\lvert00\cdots0\rangle, 
\label{eq:intermediate-state}
\end{equation}
with $n=1,\ldots,N-1$. The two endpoints are $\lvert\Phi_0\rangle=\lvert\widetilde{\psi}\rangle$ and
$\lvert\Phi_N\rangle=\lvert\psi_{\mathrm{target}}\rangle$.
Defining
\begin{equation}
d_{j,k}=d_{\mathrm{FS}}\!\left(\lvert\Phi_j\rangle,\lvert\Phi_k\rangle\right),
\end{equation}
repeated application of the triangle inequality gives
\begin{equation}
d_{N,0}\leq d_{N,N-1}+d_{N-1,0}\leq\cdots\leq\sum_{n=1}^{N}d_{n,n-1}.
\label{eq:fs-chain}
\end{equation}
Since all $U^{[n]}$ and $\widetilde U^{[n]}$ are unitary, and $F_n=1-\mathcal{L}_n$ with $F_n,\mathcal{L}_n\in[0,1]$, we have
\begin{equation}
d_{n,n-1}=\arccos\sqrt{F_n}=\arcsin\sqrt{\mathcal{L}_n}.
\label{eq:local-fs-distance}
\end{equation}
Likewise,
\begin{equation}
d_{N,0}=\arccos\sqrt{F}=\arcsin\sqrt{\mathcal{L}}.
\label{eq:global-fs-distance}
\end{equation}
Combining Eqs.~\eqref{eq:fs-chain}--\eqref{eq:global-fs-distance} yields
\begin{equation}
\arcsin\sqrt{\mathcal{L}}\leq\sum_{n=1}^{N}\arcsin\sqrt{\mathcal{L}_n}.
\label{eq:global-local-bound}
\end{equation}
Equivalently,
\begin{equation}
\mathcal{L}\leq
\sin^2\left[
\min\left\{
\frac{\pi}{2},
\sum_{n=1}^{N}\arcsin\sqrt{\mathcal{L}_n}
\right\}
\right].
\label{eq:global-infidelity-bound}
\end{equation}
Therefore, a sufficient condition for the global infidelity to satisfy $\mathcal{L}\leq\mathcal{\epsilon}_{\mathrm{global}}$ is
\begin{equation}
\sum_{n=1}^{N}\arcsin\sqrt{\mathcal{L}_n}\leq\arcsin\sqrt{\mathcal{\epsilon}_{\mathrm{global}}}.
\label{eq:target-condition}
\end{equation}

In the small-error regime, $\mathcal{L}_n\ll1$, we have $\arcsin\sqrt{\mathcal{L}_n}\simeq\sqrt{\mathcal{L}_n}$. Equation~\eqref{eq:global-local-bound} then gives
\begin{equation}
\mathcal{L}\lesssim\left(\sum_{n=1}^{N}\sqrt{\mathcal{L}_n}\right)^2.
\label{eq:small-error-bound}
\end{equation}
In particular, if $\mathcal{L}_n\leq\mathcal{\epsilon}_{\mathrm{local}}$ for all $n$, a sufficient condition for $\mathcal{L}\leq\mathcal{\epsilon}_{\mathrm{global}}$ is
\begin{equation}
\mathcal{\epsilon}_{\mathrm{local}}\leq
\sin^2\left(
\frac{\arcsin\sqrt{\mathcal{\epsilon}_{\mathrm{global}}}}{N}
\right)
\simeq\frac{\mathcal{\epsilon}_{\mathrm{global}}}{N^2},
\label{eq:local-threshold}
\end{equation}
where the final approximation applies for $\mathcal{\epsilon}_{\mathrm{global}}\ll1$.

This bound is conservative. This important result demonstrates that in DCE, in order to reach a global error $\mathcal{\epsilon}_{\mathrm{global}}$, the local error $\mathcal{\epsilon}_{\mathrm{local}}$
does not need to decrease exponentially with respect to $N$. 

\section*{Acknowledgment}
This work was supported by the Quantum Science and Technology-National Science and Technology Major Project (2023ZD0300200) and the Fundamental Research Funds for the Central Universities.

\begin{suppinfo}
	
    The Supporting Information is available as an ancillary file with this submission.
	
	\begin{quote}
		Detailed MPS-compilation methods; Convergence of MPD+EV infidelity evaluation with respect to the cutoff for bond dimension; Effects of optimizers and initialization in DCD and DCE.
	\end{quote}
	
	\end{suppinfo}

\section*{Data and Software Availability}
Source code to reproduce the reported results can be found
at Ref. \cite{FOCUS-MPS2Circuit}.
Examples can be found at Ref. \cite{MPS2Circuit-data}.

\bibliography{main}

\end{document}